\documentclass[epj]{svjour}
\usepackage{graphics}
\usepackage{threeparttable}
\usepackage{color}

\begin{document}

\title{Quasiparticle dynamics in ferromagnetic compounds of the Co-Fe and Ni-Fe systems}
\author{I.~A. Nechaev\inst{1,2} \and E.~V. Chulkov\inst{3,4,5}}

\institute{Department of Theoretical Physics, Nekrasov Kostroma State University, 156961 Kostroma, Russia
\and Research-Education Center ``Physics and Chemistry of High-Energy Systems'', Tomsk State University,
634050 Tomsk, Russia \and Donostia International Physics Center (DIPC), P. de Manuel Lardizabal, 4, 20018,
San Sebasti{\'a}n, Basque Country, Spain \and Departamento de F{\'\i}sica de Materiales, Facultad de Ciencias
Qu{\'\i}micas, UPV/EHU, Apdo. 1072, 20080 San Sebasti\'an, Basque Country, Spain \and Centro de F\'isica de
Materiales CFM - Materials Physics Center MPC, Centro Mixto CSIC-UPV/EHU, 20018 San Sebasti\'an, Spain}
\date{Received: date / Revised version: date}
\abstract{We report a theoretical study of the quasiparticle lifetime and the quasiparticle mean free path
caused by inelastic electron-electron scattering in ferromagnetic compounds of the Co-Fe and Ni-Fe systems.
The study is based on spin-polarized calculations, which are performed within the $G^0W^0$ approximation for
equiatomic and Co(Ni)-rich compounds, as well as for their constituents. We mainly focus on the spin
asymmetry of the quasiparticle properties, which leads to the spin-filtering effect experimentally observed
in spin-dependent transport of hot electrons and holes in the systems under study. By comparing with
available experimental data on the attenuation length, we estimate the contribution of the inelastic mean
free path to this length.
}

\maketitle

\section{Introduction}\label{intro}

Spin-dependent transport of hot electrons and holes in ferromagnetic materials is one of the basic phenomena
exploited in constructing spintronic devices. As an example, one can mention spin-valve and magnetic tunnel
transistors, which are also used to experimentally study the spin-dependent transport in different
many-electron systems \cite{SVT_1,SVT_2,vanDijken,MTT}. Such a study aims to solve the problem related with
controllable search for desirable parameters of constructed devices. From theoretical side, it is important
to determine a role played by inelastic electron-electron ($e-e$) scattering in formation of operating
characteristics of the devices. To this end, the finite lifetime caused by the $e-e$ scattering and the
corresponding mean free path of electrons and holes should be analyzed as a function of exciting energy.

For pure ferromagnetic metals, a similar analysis of the mentioned quasiparticle properties was made within
both semiempirical approaches and \textit{ab initio} calculations. Some of the calculations were performed
within the $G^0W^0$ approximation (for a recent review, see, e.g., Ref. \cite{UFN_2009}). This approximation
is a non-self-consistent variant of the $GW$ approach based on Kohn-Sham states evaluated within the local
spin density approximation (LSDA) with or without a Hubbard $U$ correction as a starting point
\cite{DFT_remark}. Also, the calculations were done within the LSDA+DMFT approach
\cite{LSDA_DMFT_2006,LSDA_DMFT_2007,LSDA_DMFT_2009}, which combines the LSDA and the dynamical mean-field
theory (DMFT). This approach contains two parameters (the averaged screened Coulomb interaction $U$ and the
exchange interaction $J$), which are defined outside the scope of the approach and can be taken from
experiment or calculated, e.g., within the constrained local density approximation or the constrained random
phase approximation (see, e.g., \cite{cLDA_cRPA,cLDA,cRPA}). In that sense, the $GW$+DMFT approach
\cite{GW_DMFT} can be considered as a parameter free one.

However, in the most cases ferromagnetic alloys Co$_x$Fe$_{1-x}$ and Ni$_{x}$Fe$_{1-x}$ with large $x$ are
used in spintronic devices. To all appearance, the first attempt to theoretically study the lifetime and the
inelastic mean free path in ferromagnetic alloys and compounds has been made by the authors in
\cite{FTT_2009_FeCo}. In the work cited, the properties of quasipartilces (electrons and holes) in the
ordered (B2) and disordered (body-centered cubic) CoFe have been calculated from first principles within the
mentioned $G^0W^0$ approximation. Also, the contribution of the localized $d$ states to the above properties
has been analyzed by considering the Coulomb interaction screened by only $d$ electrons instead of the fully
screened Coulomb interaction.

In the present paper, we proceed with the theoretical study of the properties of quasiparticles in
ferromagnetic compounds, as well as in their constituents. In order to be closer to the Co- and Ni-rich
ferromagnetic alloys used in practice, among the compounds we consider Co$_3$Fe and Ni$_3$Fe with the DO$_3$
and L1$_2$ structure, respectively. We perform \textit{ab initio} calculations within the $G^0W^0$
approximation with the fully screened Coulomb interaction found in the random phase approximation (RPA). To
demonstrate possible effects that structure and stoichiometry changes may have on the quasiparticle
properties under study, we additionally consider the ferromagnetic compound NiFe with the L1$_0$ structure.
For completeness of the picture of such an effect, we also analyze the results obtained in
\cite{FTT_2009_FeCo} for the B2 CoFe.

The paper is organized as follows. In Sec.~\ref{Approx}, we briefly describe the scheme that is used to
evaluate the spin-resolved quasiparticle lifetime and inelastic mean free path within the $G^0W^0$
approximation. Also, we put here some calculation details. In Sec.~\ref{RnD}, we present our main results of
extensive calculations carried out for quasiparticle properties in the mentioned compounds and pure metals.
On the base of these results, we analyze how changes in structure and stoichiometry modify quasiparticle
properties evaluated from the $G^0W^0$ calculations. Finally, the conclusions are given in
Sec.~\ref{Conclusions}.

\section{Approximations and calculation details}\label{Approx}

In this section, to make the paper self-sustained we briefly describe the $G^0W^0$ approximation
\cite{Hedin_PR_1965} to the quasiparticle self-energy, which is used in the paper. Unless stated otherwise,
atomic units are used throughout, i.e., $e^2=\hbar=m=1$.

The $G^0W^0$ self-energy of a quasiparticle with the spin $\sigma$ is defined as
\begin{eqnarray}\label{Sigma_GW}
\Sigma_{\sigma}(\mathbf{r}_1,\mathbf{r}_2;\omega)&=&\frac{i}{2\pi}\int d\omega'
e^{i\eta\omega'}G^0_{\sigma}(\mathbf{r}_1,\mathbf{r}_2;\omega')\nonumber\\
&\times& W^0(\mathbf{r}_1,\mathbf{r}_2;\omega-\omega'),
\end{eqnarray}
where the convergence factor $\exp(i\eta\omega')$ indicates that the integration contour is closed in the
upper half-plane of $\omega'$. The Green function entering Eq.~(\ref{Sigma_GW}) is defined as \cite{Ary_BOOK}
\begin{eqnarray}\label{Green_f}
G^0_{\sigma}(\mathbf{r}_1,\mathbf{r}_2;\omega)&=&\sum_{\mathbf{k}n}^{occ}
\frac{\psi_{\mathbf{k}n\sigma}(\mathbf{r}_1)\psi^{\ast}_{\mathbf{k}n\sigma}(\mathbf{r}_2)}
{\omega-\epsilon_{\mathbf{k}n\sigma}-i\delta}\nonumber\\
&+&\sum_{\mathbf{k}n}^{unocc}
\frac{\psi_{\mathbf{k}n\sigma}(\mathbf{r}_1)\psi^{\ast}_{\mathbf{k}n\sigma}(\mathbf{r}_2)}
{\omega-\epsilon_{\mathbf{k}n\sigma}+i\delta}
\end{eqnarray}
and corresponds to the Kohn-Sham equation with the exchange-correlation potential $V_{\sigma}^{XC}$ obtained
within the LSDA. In Eq.~(\ref{Green_f}), the positive infinitesimal $\delta$ characterizes the way of going
around poles in the course of integration. The states
$\left\{\psi_{\mathbf{k}n\sigma},\epsilon_{\mathbf{k}n\sigma}\right\}$ are calculated by the self-consistent
tight-binding linear muffin-tin orbital method \cite{Andersen} with the use of the crystal potential
constructed within the atomic-sphere approximation.

The screened Coulomb interaction $W^0$ participating in the definition of the self-energy (\ref{Sigma_GW}) is
found within the RPA, where the irreducible polarizability $P^0$ is expressed as
\begin{eqnarray}\label{P0_RPA}
P^0(\mathbf{r}_1,\mathbf{r}_2;\omega)&=&\frac{i}{2\pi}\sum_{\sigma}\int
d\omega'G^0_{\sigma}(\mathbf{r}_1,\mathbf{r}_2;\omega')\nonumber
\\
&\times&G^0_{\sigma}(\mathbf{r}_1,\mathbf{r}_2;\omega'+\omega).
\end{eqnarray}
Owing to the translation symmetry, the irreducible polarizability can be expanded into a series
\cite{Ary_BOOK,GV_eliquid}
\begin{equation}\label{P0_Series}
P^0(\mathbf{r}_1,\mathbf{r}_2;\omega)=\sum_{\mathbf{k}ij}
B_{\mathbf{k}i}(\mathbf{r}_1)P^0_{ij}(\mathbf{k},\omega) B^{\ast}_{\mathbf{k}j}(\mathbf{r}_2),
\end{equation}
where $\left\{B_{\mathbf{k}i}\right\}$ is a set of basis functions, which satisfy the Bloch theorem and are
normalized to unity in the unit cell volume $\Omega$. These basis functions are constructed using pair
products of linear muffin-tin orbitals localized at the same lattice site \cite{Aryasetiawan_PRB_1994}. The
irreducible polarizability $P^0_{ij}(\mathbf{k},\omega)$ is derived with the use of the spectral function
representation. The corresponding spectral function
\begin{eqnarray}\label{S0}
S^0_{ij}(\mathbf{q},\omega)&=&\sum_{\mathbf{k}\sigma}\sum_{nn'}\left(f_{\mathbf{k}+\mathbf{q}n'\sigma} -
f_{\mathbf{k}n\sigma}\right) \langle B_{\mathbf{q}i}\psi_{\mathbf{k}n\sigma}|
\psi_{\mathbf{k}+\mathbf{q}n'\sigma}\rangle \nonumber \\
&\times& \langle \psi_{\mathbf{k}+\mathbf{q}n'\sigma} | \psi_{\mathbf{k}n\sigma}B_{\mathbf{q}j}\rangle
\delta[\omega-(\epsilon_{\mathbf{k}+\mathbf{q}n'\sigma}-\epsilon_{\mathbf{k}n\sigma})],
\end{eqnarray}
where $f_{\mathbf{k}n\sigma}$ is the Fermi factor, is obtained by substituting the Green function of
Eq.~(\ref{Green_f}) into the expression (\ref{P0_RPA}) for $P^0$ and using the expansion (\ref{P0_Series})
along with the relation $\mathrm{Im}P^0_{ij}(\mathbf{k},\omega)=-\pi S_{ij}^0(\mathbf{k},\omega)
\mathrm{sgn}(\omega)$. In calculations, the $\delta$-function is replaced by the Gaussian
$\exp(-\omega^2/\gamma^2)/(\gamma\sqrt{\pi})$ with $\gamma=0.136$ eV \cite{Zhukov_PRB_2001}.

The inverse quasiparticle lifetime or decay rate is defined as $\tau_{\mathbf{k}n\sigma}^{-1}
=2Z_{\mathbf{k}n\sigma}| \mathrm{Im} \Sigma_{\mathbf{k}n\sigma} (\epsilon_{\mathbf{k}n\sigma})|$, where the
renormalization factor $Z_{\mathbf{k}n\sigma}$ is given by
$$
Z_{\mathbf{k}n\sigma}=\left[1-\frac{\partial\mathrm{Re}\Sigma_{\mathbf{k}n\sigma}(\omega)} {\partial\omega}
\right]^{-1}_{\omega=\epsilon_{\mathbf{k}n\sigma}}.
$$
With the use of the expressions presented above, the imaginary part of the self-energy matrix elements
$\Sigma_{\mathbf{k}n\sigma}(\omega) = \langle\psi_{\mathbf{k}n\sigma}| \Sigma_{\sigma}(\omega)
|\psi_{\mathbf{k}n\sigma}\rangle$ entering the definition of the quasiparticle decay rate takes the form
\begin{eqnarray}\label{ImSigma_GW}
\mathrm{Im}\Sigma_{\mathbf{k}n\sigma}(\omega)&=&\mp\sum_{\mathbf{q}n'}\sum_{ij}
\langle\psi_{\mathbf{k}n\sigma}\psi_{\mathbf{q}-\mathbf{k}n'\sigma}|B_{\mathbf{q}i}\rangle \nonumber\\
&\times&\mathrm{Im}W^0_{ij}(\mathbf{q},\pm\epsilon_{\mathbf{q}-\mathbf{k}n'\sigma}\mp\omega)  \\
&\times&\langle B_{\mathbf{q}j}|\psi_{\mathbf{q}-\mathbf{k}n'\sigma}\psi_{\mathbf{k}n\sigma}
\rangle\Theta(\pm\epsilon_{\mathbf{q}-\mathbf{k}n'\sigma}\mp\omega) ,\nonumber
\end{eqnarray}
where the upper (lower) sign corresponds to the exciting energy $\omega\leq E_F$ ($\omega>E_F$) and the sum
over occupied (unoccupied) states. Here, $E_F$ is the Fermi energy. In Eq.~(\ref{ImSigma_GW}), $\Theta(x)$ is
the step function and $W_{ij}$ are matrix elements of the screened interaction in the
$\left\{B_{\mathbf{k}i}\right\}$ basis, which are defined by $P^0_{ij}(\mathbf{k},\omega)$ and
Coulomb-interaction matrix elements \cite{Ary_BOOK}. Having obtained the imaginary part of
$\Sigma_{\mathbf{k}n\sigma}(\omega)$, the real part of the latter is found from the Hilbert transform.

Both in Eq.~(\ref{ImSigma_GW}) and in Eq.~(\ref{S0}), the $s$, $p$, and $d$ bands are involved in the sums
over the band indices. As to the sums over momenta $\mathbf{k}$, a set of points $20\times20\times20$ is used
for the face- and body-centered cubic (fcc and bcc) structures and the B2, L1$_{2}$, and DO$_{3}$ structures.
For the L1$_{0}$ structure with the primitive vectors ($-1/2,1/2,0$), ($1/2,1/2,0$), and ($0,0,1$), we use a
set of points $20\times20\times12$. The quasiparticle lifetimes are calculated at all $\mathbf{k}$-points of
these sets, which belong to the irreducible Brillouin zone, what ensures the momentum averaging of the
lifetimes to be well converged. The number of optimal product basis functions per atom is 40.

As regards the well-known problem with the slow convergence of $\mathrm{Re}\Sigma_{\mathbf{k}n\sigma}$ with
respect to the number of unoccupied states and basis functions, it is worth noting that in order to consider
$G^0W^0$ corrections to the LSDA band structure, a minimal requirement in our case is to involve $f$ band and
to increase the number of optimal product basis functions per atom to $115$ with the inclusion of core states
(see, e.g., \cite{Ary_BOOK,Aryasetiawan_1994_1998}). Since we are interested in the quasiparticle lifetime
determined by the self-energy, where $W^0$ is computed within the RPA instead of, e.g., the plasmon-pole
model, the mentioned modifications of the calculation parameters make a study of the lifetime in the
ferromagnetic compounds hardly realized. At that, as an analysis has shown by the example of Ni,
$\mathrm{Im}\Sigma_{\mathbf{k}n\sigma}$ and $Z_{\mathbf{k}n\sigma}$ converges noticeably faster than
$\mathrm{Re}\Sigma_{\mathbf{k}n\sigma}$. At the calculation parameters used in the paper, the imaginary part
of the self-energy and the renormalization factor do not differ substantially (within $4-7\%$ on average in
the considered energy range) from these obtained within the requirement just mentioned. At least, the found
differences do not affect the issues, which we discuss below.

In order to estimate the inelastic mean free path (IMFP) of quasiparticles as
$\lambda^{e-e}_{\mathbf{k}n\sigma}=|\mathbf{v}_{\mathbf{k}n\sigma}|\tau_{\mathbf{k}n\sigma}$, the
quasiparticle velocity $\mathbf{v}_{\mathbf{k}n\sigma}$ can be evaluated from the expression (see, e.g., Ref.
\cite{GV_eliquid})
\begin{equation}\label{velocity}
\mathbf{v}_{\mathbf{k}n\sigma}=Z_{\mathbf{k}n\sigma}\left(\mathbf{v}^0_{\mathbf{k}n\sigma}
+\nabla_{\mathbf{k}}\mathrm{Re} \Delta\Sigma_{\mathbf{k}n\sigma}(\omega)
|_{\omega=\epsilon_{\mathbf{k}n\sigma}}\right),
\end{equation}
where $\mathbf{v}^0_{\mathbf{k}n\sigma}=\nabla_{\mathbf{k}}\epsilon_{\mathbf{k}n\sigma}$ and the difference
$\Delta\Sigma_{\mathbf{k}n\sigma}(\omega) = \Sigma_{\mathbf{k}n\sigma}(\omega) -
\langle\psi_{\mathbf{k}n\sigma}|V_{\sigma}^{XC}|\psi_{\mathbf{k}n\sigma}\rangle$, which is treated as an
expansion parameter. In the present calculations, for simplicity we neglect the second term in the right hand
side of Eq.~(\ref{velocity}). This leads to the IMFP given by the frequently used formula
\begin{equation}\label{IMFP_s}
\lambda^{e-e}_{\mathbf{k}n\sigma}=\left|\frac{\mathbf{v}^0_{\mathbf{k}n\sigma}}
{2\mathrm{Im}\Sigma_{\mathbf{k}n\sigma}(\epsilon_{\mathbf{k}n\sigma})}\right|.
\end{equation}
It is worth noting that this formula, which, e.g., in \cite{Silkin_PRB_2003} has allowed to obtain
theoretical results in good agreement with experimental data in the case of beryllium, does not contain the
real part of the self-energy matrix elements. On the one hand, this means that many-body effects cannot be
taken into account in full measure. But on the other hand, such an IMFP is free of the convergence problem
mentioned above and is completely defined by the underlying (LSDA, in our case) band structure. For a
pragmatic point of view, the used approximation seems to be a quite accurate and feasible method for
predicting the quasiparticle lifetime and the IMFP in ferromagnetic metals and compounds on the same
footings.

\section{Results and discussion}\label{RnD}

In the present calculations, the cubic lattice parameter $a$ for CoFe (B2) and Co$_3$Fe (DO$_3$) is chosen to
be equal to 2.856 \AA \cite{Numerical_Data} and 5.657 \AA \cite{Kim_Lee_Park}, respectively. The same lattice
parameter as for CoFe is used for the bcc Fe and the bcc Co. This parameter is very close both to the
equilibrium parameter of pure iron and to the parameter of bcc cobalt films ($2.82\pm0.01$ \AA
\cite{Co_on_GaAs}) grown on GaAs. Moreover, the parameter $a$ of Co$_3$Fe with the bcc-based DO$_3$ structure
is about doubled $a$ of the mentioned bcc Co films. In NiFe (L1$_0$), the parameter $a$ has the value of
3.585 \AA, which was estimated by linear interpolation between 57.0 and 44.3 at. \% Ni in Ni$_{1-x}$Fe$_x$
fcc alloys \cite{Numerical_Data}. For Ni$_3$Fe with the L1$_2$ structure and for the fcc Ni, the lattice
parameter is taken to be equal to 3.552 \AA \cite{Numerical_Data}.

\subsection{Co-Fe system}\label{RnD:Co-Fe}

\begin{figure}[tbp]
\centering \resizebox{0.45\textwidth}{!}{
\includegraphics{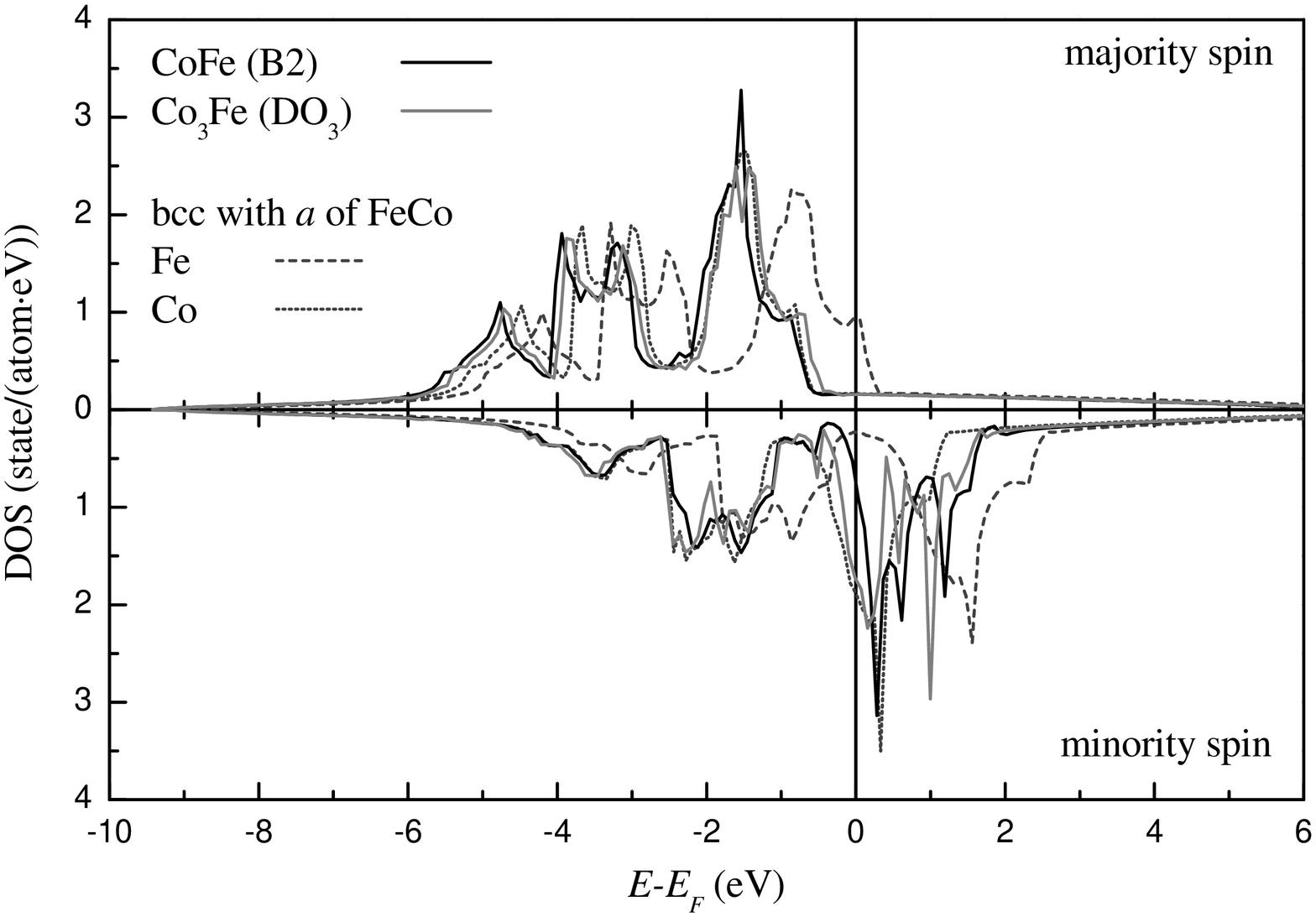}}
  \caption{Density of states in the ferromagnetic compounds CoFe and Co$_3$Fe with the B2 and DO$_3$ structure, respectively,
  as well as in the bcc Fe and the bcc Co with the lattice parameter of CoFe.} \label{fig:1}
\end{figure}

Fig. \ref{fig:1} shows the calculated density of states (DOS) both in the considered ferromagnetic compounds
of the Co-Fe system and in the pure metals forming these compounds. The presented curves demonstrate the
behavior of the DOS as a function of energy, which corresponds to the LSDA band structure. We plot these
curves with the purpose to explore specific features, which already at the stage of analyzing the DOS allow
one to qualitatively describe the behavior of the ratio $\tau_{\uparrow}/\tau_{\downarrow}$ between the
lifetimes of quasiparticles with different spin orientations and how this ratio changes as we move from Fe,
to CoFe and Co$_3$Fe, and then to Co. Such a description is based on the fact that the DOS directly depends
on dispersion of $\epsilon_{\mathbf{k}n}$ that along with $\psi_{\mathbf{k}n}$ determines quasiparticle
lifetimes.

Due to the bcc-based structure of the considered materials, the DOS has a form that is typical for the bcc
transition metals: a profound minimum halves the $d$-band \cite{Papaconst_BOOK}. Moreover, the DOS in the bcc
Fe is similar to that in the bcc Co, where the main effect caused by moving from iron to cobalt is a change
in $E_F$ due to an additional electron. As a consequence, the densities of states in Co, CoFe, and Co$_3$Fe
are very close to each other, especially in the spin-up subsystem. As to the spin-down subsystem, a
distinguishing feature of these cobalt-containing materials is the DOS shape in the vicinity (approximately
$\pm1$ eV) of the Fermi level. In this sense, an important difference between, e.g., Co$_3$Fe and CoFe is
that the former has the DOS (predominantly formed by $d$ states) that is higher than that in the latter. This
means that in the vicinity of $E_F$ the DO$_3$ Co$_3$Fe is more spin polarized than the B2 CoFe.

The obtained LSDA band structure is characterized by the local magnetic moments, which are listed for
comparison with experimental and other theoretical data in Table \ref{table:1}. Note that in the case of the
DO$_3$ Co$_3$Fe two values of $\mu_{\mathrm{Co}}$ are presented. It caused by the fact that there are two
inequivalent sites for Co atoms in the DO$_3$ structure. As is seen from the table, in the considered
compounds the magnetic moment for Co remains practically unchanged as compared with that in the pure bcc
cobalt. At the same time, the magnetic moment for Fe becomes noticeably larger upon moving from the pure bcc
iron to the compounds. Such a behavior of $\mu_{\mathrm{Co}}$ and $\mu_{\mathrm{Fe}}$ is in agreement with
available experimental data.

\begin{table}
\caption{Calculated local magnetic moments (in $\mu_B$)} \label{table:1} \centering
\begin{tabular}{lccl}
\hline\noalign{\smallskip}
Material & $\mu_{\mathrm{Fe}}$ & $\mu_{\mathrm{Co}}$ & \\
\noalign{\smallskip}\hline\noalign{\smallskip}
 Co (bcc)           &      & 1.72           & This work                          \\
                    &      & 1.3-1.7        & Exp.~\cite{Hiper_field,Thermal_magnons,Magn_prop}        \\
 Co$_3$Fe (DO$_3$)  & 2.62 & 1.72 / 1.71    & This work                          \\
                    & 2.61 & 1.78 / 1.78    & Theory~\cite{Kumar_EPJB_2010}        \\
 CoFe (B2)          & 2.73 & 1.74           & This work                          \\
                    & 2.82 & 1.74           & Theory~\cite{Pourovskii_PRB_2005}    \\
                    & 2.92$\pm$0.02 & 1.62$\pm$0.02           & Exp.~\cite{Spin_density}    \\
 Fe (bcc)           & 2.21 &                & This work                          \\
                    & 2.22 &                & Exp.~\cite{Danan}                  \\
\noalign{\smallskip}\hline
\end{tabular}
\end{table}

As noted above, the mentioned changes in filling, width, and shape of $d$-band should affect the lifetime as
a function of exciting energy. In the case of paramagnetic simple, noble, and transition metals, such an
influence has been considered in \cite{Campillo_2000,Zhukov_2002,GW_param} in details. In the present work,
we can additionally analyze the ratio $\tau_{\uparrow}/\tau_{\downarrow}$. Actually, as is seen from
Fig.~\ref{fig:1}, in the vicinity of $E_F$ (especially for positive exciting energies) the density of spin-up
states is considerably smaller than the density of spin-down states. As a result, we can expect that, at
least in this energy rage, the ratio $\tau_{\uparrow}/\tau_{\downarrow}$ will be substantially greater than
unity. An exception is constituted by the bcc iron that should be characterized by the above ratio being
smaller than unity.

\begin{figure}[tbp]
\centering \resizebox{0.45\textwidth}{!}{
\includegraphics{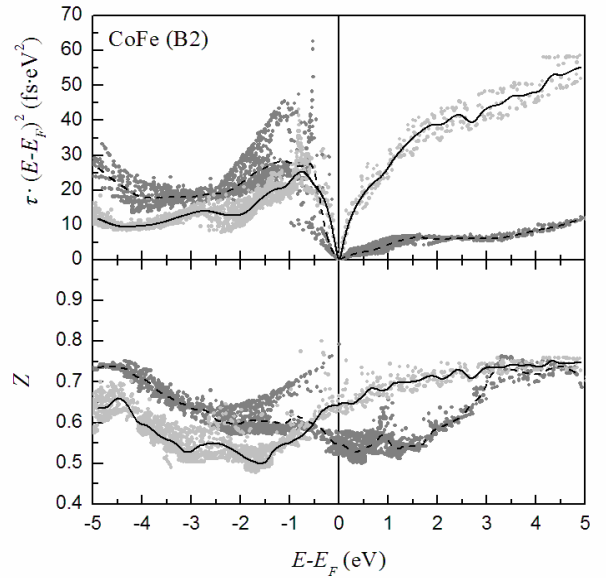}}
  \caption{Scaled lifetime $\tau_{\mathbf{k}n\sigma}\times(\epsilon_{\mathbf{k}n\sigma}-E_F)^2$ (upper panel) and
  renormalization factor $Z_{\mathbf{k}n\sigma}$ (lower panel) as functions of the exciting energy
  $\epsilon_{\mathbf{k}n\sigma}-E_F$ for spin-up (light gray points) and spin-down (dark gray points) quasiparticles
  in the B2 CoFe ferromagnetic compound. Solid and dashed lines show the corresponding scaled lifetimes
  averaged over momentum $\mathbf{k}$ for a given exciting energy.} \label{fig:2}
\end{figure}

\begin{figure}[tbp]
\centering \resizebox{0.45\textwidth}{!}{
\includegraphics{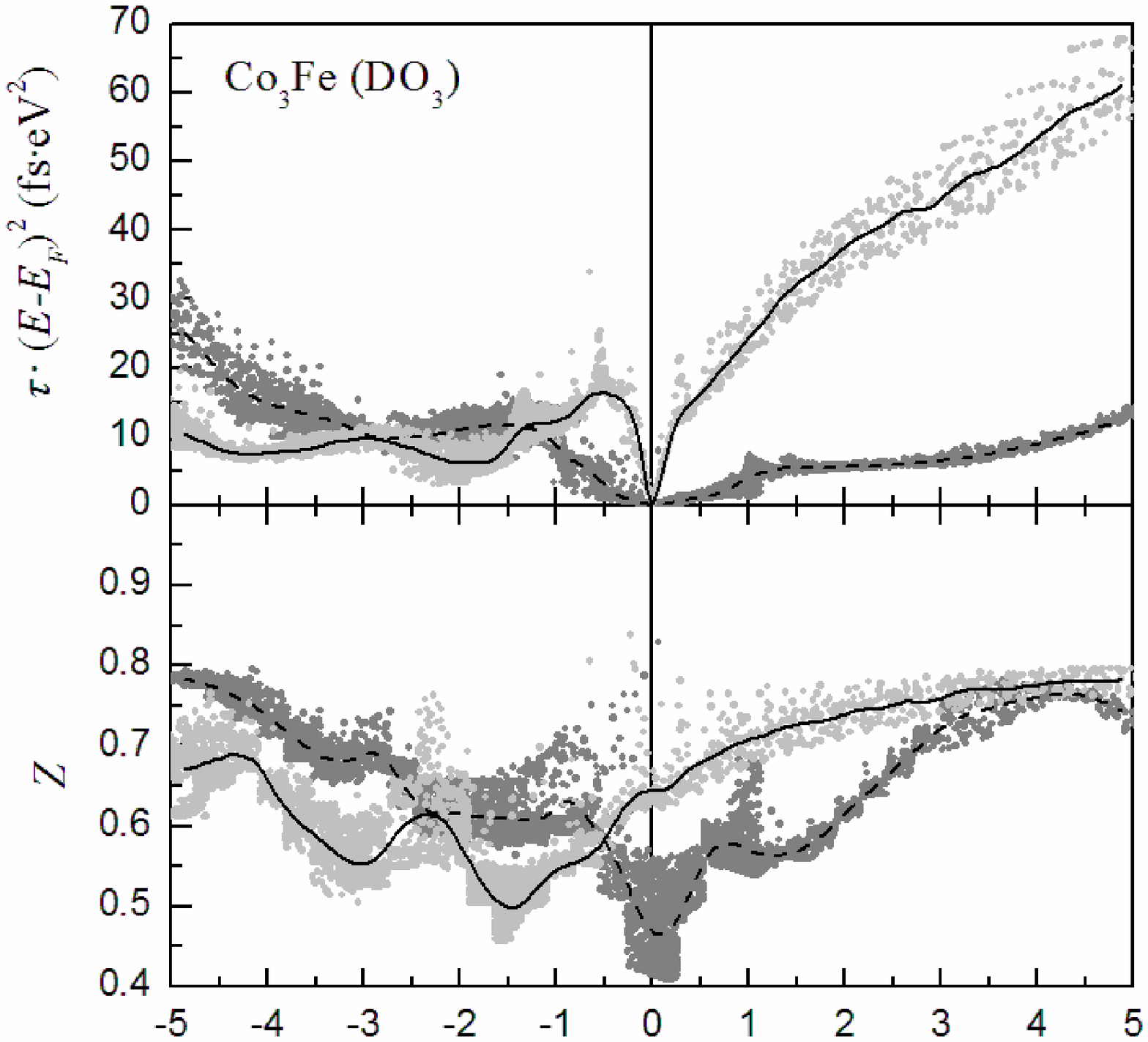}}
  \caption{Same as in Fig.~\ref{fig:2}, but in the DO$_3$ Co$_3$Fe ferromagnetic compound.} \label{fig:3}
\end{figure}

The results of our calculations of the quasiparticle properties in the B2 CoFe are shown in Fig.~\ref{fig:2}.
In this figure, we demonstrate both momentum-resolved and momentum-averaged lifetime and renormalization
factor for spin-up and spin-down quasiparticles. In order to clearly represent the exciting-energy dependence
of the quasiparticle lifetime, the so-called scaled lifetime (that is the lifetime multiplied by squared
exciting energy) is shown. Analyzing the data presented in Fig.~\ref{fig:2}, we would like to note the
significant difference between the lifetime of electrons with spin up and spin down. Such a difference is one
of the main reason of appearing the so-called spin-filtering effect that is observed in spin-dependent
transport of electrons in ferromagnetic materials \cite{vanDijken,UFN_2009}. The lifetime of holes
(especially with spin down) depends strongly on the momentum $\mathbf{k}$ at a given exciting energy. It
follows from the observed spread in values of $\tau_{\mathbf{k}n\sigma}$ at
$\epsilon_{\mathbf{k}n\sigma}<E_F$.

As regards the renormalization factor characterizing the spectral weight of quasiparticles, like the
densities of states for corresponding spin subsystems the factors $Z_{\mathbf{k}n\uparrow}$ and
$Z_{\mathbf{k}n\downarrow}$ as functions of exciting energy are quite similar, in many respects, but with
some shift on energy scale, which is determined by the exchange splitting of the bands. For both spins, there
is a ``depression'' in the energy range that corresponds to the dominant contribution of the localized $d$
states to the total DOS. This means that many-body effects have the most impact on these states (see, e.g.,
Ref.~\cite{Zein_2002}). In the energy range from $\sim$0 to $\sim$3 eV, where a comparatively high density of
spin-down states corresponds to a low density of spin-up states, one observes the largest difference between
$Z_{\mathbf{k}n\uparrow}$ and $Z_{\mathbf{k}n\downarrow}$. At higher energies, the renormalization factor of
electrons depends weakly on spin.

Fig.~\ref{fig:3} shows our results on the quasiparticle lifetime and renormalization factor in the case of
the ferromagnetic compound Co$_3$Fe with the DO$_3$ structure. As a whole, for electrons the scaled lifetime
(momentum-averaged and momentum-resolved) differs slightly from that in the B2 CoFe, whereas for holes in the
exciting-energy range from 0 to -3 eV the quantity under consideration demonstrates noticeably smaller values
and substantially weaker dependence on the momentum $\mathbf{k}$ at a given exciting energy. At that, in the
vicinity of the Fermi energy, the scaled lifetime of spin-up holes is longer than that of spin-down holes
practically to the same extent as the scaled lifetime of spin-up and spin-down electrons.

Upon moving from the B2 CoFe to the DO$_3$ Co$_3$Fe, the renormalization factor undergoes a visible change in
momentum dependence at a given energy (especially at $E_F$) and, being momentum-averaged, in dependence on
exciting energy. At that, the mentioned correlation between values of the renormalization factor and the
$d$-states contribution to the DOS remains evident as before. Note that in this case
$Z_{\mathbf{k}n\downarrow}$ has a profound minimum at the Fermi level. This minimum is caused by a quite
narrow and high peak located practically at $E_F$ in the density of $d$ states. The factor
$Z_{\mathbf{k}n\uparrow}$ demonstrates a similar minimum at about $-1.5$ eV, which is also correlated with a
peak in the density of $d$ states (see Fig.~\ref{fig:1}).

\begin{figure}[tbp]
\centering \resizebox{0.45\textwidth}{!}{
\includegraphics{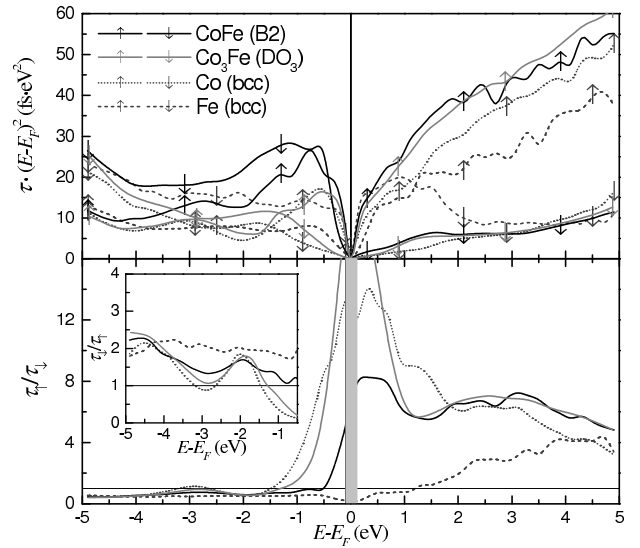}}
  \caption{Scaled lifetime averaged over momentum $\mathbf{k}$ for a given exciting energy (upper panel) and ratio
  $\tau_{\uparrow}/\tau_{\downarrow}$ between the averaged lifetimes of quasiparticles with different spin
  orientations (lower panel) as functions of exciting energy
  for the ferromagnetic compounds CoFe and Co$_3$Fe with the B2 and DO$_3$ structures, respectively,
  as well as for the bcc Fe and the bcc Co with the lattice parameter of CoFe. Inset in the lower panel shows
  the inverse ratio $\tau_{\downarrow}/\tau_{\uparrow}$ for holes.} \label{fig:4}
\end{figure}

\begin{figure}[tbp]
\centering \resizebox{0.45\textwidth}{!}{
\includegraphics{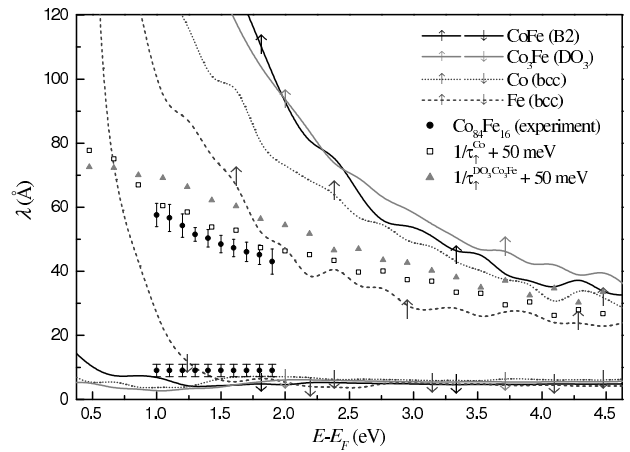}}
  \caption{Dependence of the momentum-averaged IMFP $\lambda^{e-e}_{\mathbf{k}n\sigma}$ on exciting energy.
  The experimental data on the attenuation length for spin-up and spin-down electrons in
  Co$_{84}$Fe$_{16}$ are taken from \cite{vanDijken}} \label{fig:5}
\end{figure}

In order to demonstrate what effect the changes in structure and stoichiometry have on the lifetime of
quasiparticle in the considered ferromagnetic materials, we show all the obtained results on the Co-Fe system
in Fig.~\ref{fig:4}. As is evident from the figure (see the upper panel), the lifetime of spin-up electrons
in the compounds is longer than that in their constituents. Spin-down electrons have the longest lifetime in
the bcc iron practically in the whole considered energy range. For holes, the longest lifetimes are observed
in the B2 CoFe in both spin subsystems. Note the closeness of the hole lifetimes in the DO$_3$ Co$_3$Fe and
in the bcc cobalt. In fact, in all the cases, we deal with a spin asymmetry of the quasiparticle properties,
which can be represented by the already mentioned ratio of the lifetimes of spin-up and spin-down
quasiparticles. A deviation of the ratio from unity reveals the spin asymmetry of the considered quantity. In
Fig.~\ref{fig:4}, we also plot the ratio $\tau_{\uparrow}/\tau_{\downarrow}$ between the spin-up and
spin-down momentum-averaged lifetimes (see the lower panel). One can see that the expected values
qualitatively estimated from the shape of the DOS in the vicinity of the Fermi level are confirmed by the
results presented in the figure. Actually, in the exciting-energy range from 0 to $\sim$0.7 eV the ratio
$\tau_{\uparrow}/\tau_{\downarrow}$ amounts to 8 for the B2 CoFe and exceeds 12 for the DO$_3$ Co$_3$Fe and
the bcc cobalt. For the bcc iron, the ratio varies from $\sim$0.2 to $\sim$0.8 in the above energy range and
only starting from $\sim$1.2 eV becomes greater than unity. However, for holes in Fe the inverse ratio
$\tau_{\downarrow}/\tau_{\uparrow}$ depends weakly on energy and comes to $\sim$2 (see the inset in the lower
panel of Fig.~\ref{fig:4}). It is worth noting that in contrast to the bcc iron and the bcc cobalt the
considered ferromagnetic compounds are characterized by the ratio $\tau_{\uparrow}/\tau_{\downarrow}$, which
over the wide energy range (from $\sim1$ to 5 eV) is large (about six) and varies slightly with exciting
energy. For electrons with energies less than $\sim1$ eV, the DO$_3$ Co$_3$Fe provides the largest
$\tau_{\uparrow}/\tau_{\downarrow}$.

As regards the IMFP expressed as the product of the lifetime and the quasiparticle velocity, an analysis has
shown (see, e.g., Ref.~\cite{FTT_2009_FeCo}) that a noticeable difference between spin-up-electron and
spin-down-electron velocities increases the revealed significant difference between the lifetime of electrons
with spin up and spin down. This means that for the ferromagnetic materials under study the calculations
within the $G^0W^0$ approximation predict a strong spin-filtering effect, which manifests itself in a giant
spin asymmetry of the IMFP as is clearly seen from Fig.~\ref{fig:5}. However, the experimental data (also
presented in Fig.~\ref{fig:5}) on the spin-dependent transport of electrons in Co$_{84}$Fe$_{16}$ films on
the GaAs(001) surface \cite{vanDijken} indicate that the mentioned effect is not so strong. At that, the
calculated momentum-averaged IMFP for spin-down electrons in the compounds and in the bcc cobalt is quite
close to the corresponding attenuation length and nicely reproduces the weak energy-dependence of the latter,
while the difference between the experimentally observed attenuation length and the calculated IMFP for
spin-up electrons is substantial.

To analyze the discrepancy between the theoretical and experimental data, first, we should note that the IMFP
calculated within the $G^0W^0$ approximation does not include the contribution of the decay channels caused
by spin fluctuations. As was shown in \cite{Zhukov_PR,NeChul,MoRh} for paramagnetic and ferromagnetic
materials, taking these fluctuations into consideration leads to a decrease in the quasiparticle lifetime
and, as a consequence, in the IMFP $\lambda^{e-e}_{\sigma}$. At that, $\lambda^{e-e}_{\uparrow}$ appears to
be most sensitive (see, e.g., Ref.~\cite{Zhukov_PR}). Next, following \cite{vanDijken} we represent the
measured attenuation length as a result of the sum of two terms:
$1/\lambda^{exp}_{\sigma}=1/\lambda^{e-e}_{\sigma}+ 1/\lambda^{extra}_{\sigma}$, where, in addition to the
inelastic electron-electron scattering contribution (the calculated $\lambda^{e-e}_{\sigma}$), there is a
contribution that includes the terms coming from quasielastic scattering by phonons and spin waves and from
elastic electron scattering by defects and impurities. The latter depends on type of defects and their number
and represents the term, of which contribution to the spin asymmetry of the attenuation length is expected to
be minimal (see, e.g., Ref.~\cite{Vescovo_PRB_1995}). The contribution of electron-magnon scattering is known
to be significant for spin-down electrons (see, e.g., \cite{Kleinman_PRB_1978}). The term caused by
electron-phonon scattering, has an effect on both spin-down and spin-up electrons. The dependence of this
term on exciting energy is almost completely determined by the electron velocity. This means that the
corresponding decay rate is nearly constant. To illustrate how the inclusion of a contribution caused by a
decay process weakly dependent on exciting energy can affect the attenuation length of spin-up electrons, we
approximate $1/\lambda^{extra}_{\uparrow}$ as $\Gamma_{\uparrow}/v_{\uparrow}$ with $\Gamma_{\uparrow}=50$
meV. In the order of magnitude, this $\Gamma_{\uparrow}$ is close, e.g., to the electron-phonon broadening
observed for spectral lines in different metals (see, e.g., \cite{NECHUL_Aluminum}). The obtained results are
presented in Fig.~\ref{fig:5}. In the case of the bcc cobalt, we have the least difference between the
experimental and theoretical data. To this, the behavior of the attenuation length as a function of exciting
energy is nicely reproduced. Thus, in order to reach a satisfactory agreement with the experimental data, at
least the electron-phonon broadening should be included.

\subsection{Ni-Fe system}\label{RnD:NiFe}

\begin{figure}[tbp]
\centering \resizebox{0.45\textwidth}{!}{
\includegraphics{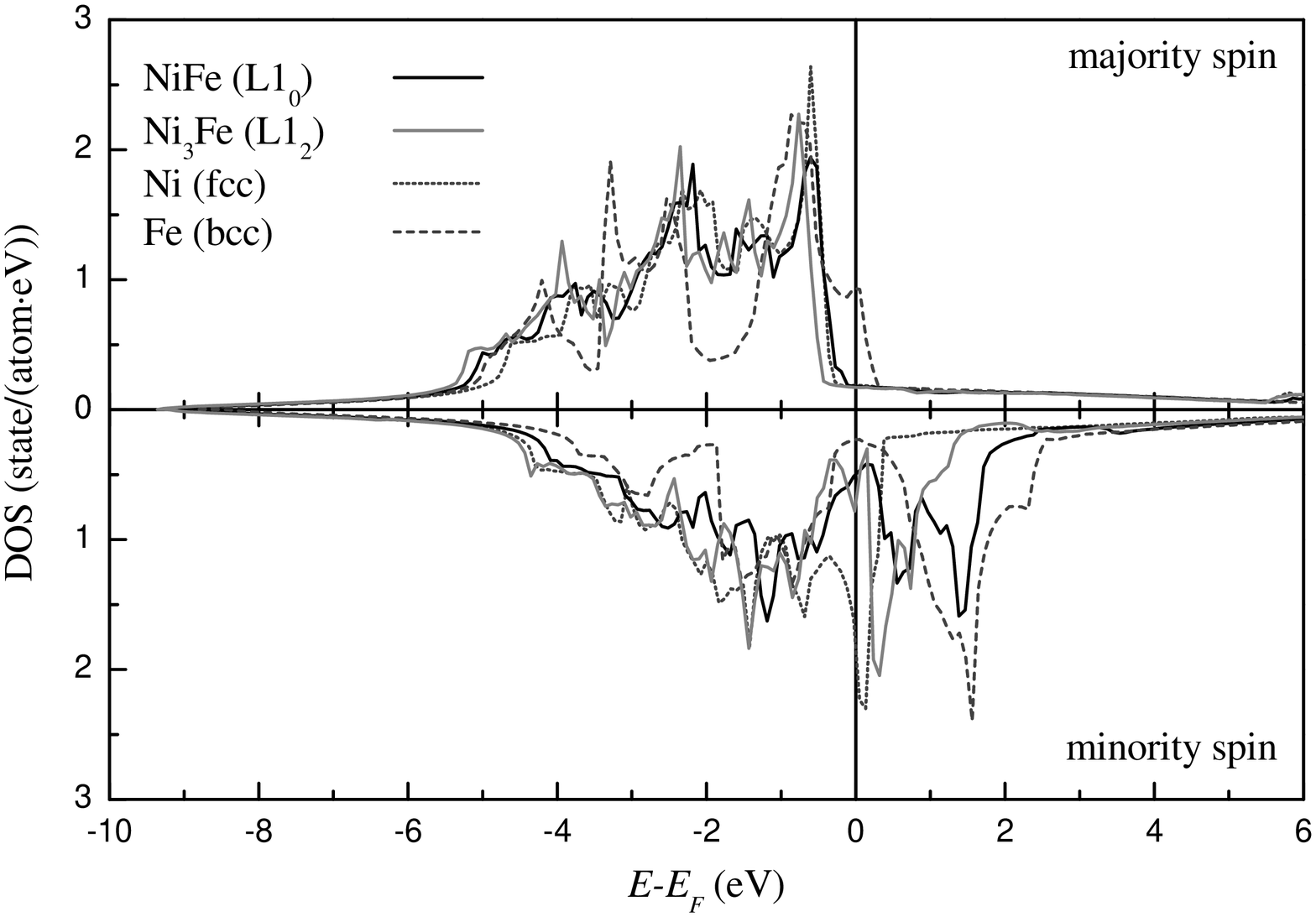}}
  \caption{Density of states in the ferromagnetic compounds NiFe and Ni$_3$Fe with the L1$_0$ and L1$_2$ structures,
  respectively, as well as in the bcc Fe and the fcc Ni.} \label{fig:6}
\end{figure}

In Fig.~\ref{fig:6}, we show the results of the DOS calculations for two compounds of the Ni-Fe system and
the pure fcc nickel within the LSDA. The fcc nickel and the compounds have the density of spin-up states with
the continuous $d$-band that is peculiar for fcc-based structures. At that the Fermi level is situated quite
close to the edge of the $d$-band. The density of spin-down states possesses the fcc-type $d$-band in the
case of fcc nickel only, where the Fermi level corresponding to a peak at the edge of the $d$-band makes this
ferromagnetic metal to be highly spin polarized. The compounds NiFe and Ni$_3$Fe demonstrate the DOS with the
minimum at $E_F$, which is rather close to the ordinary half-filled bcc $d$-band (especially in the L1$_0$
NiFe case). This means that at least in the vicinity of  $E_F$ the fcc nickel should demonstrate a quite
large ratio $\tau_{\uparrow}/\tau_{\downarrow}$ as compared with the compounds.

The found local magnetic moments are presented in Table \ref{table:2}. As is evident from the table, again,
as in the case of the Co-Fe system, in compounds the magnetic moment for Fe becomes larger than in pure iron.
At that, the largest moment $\mu_{\mathrm{Fe}}$ is observed in the L1$_2$ Ni$_3$Fe. As to Ni atoms, with good
agreement with experimental data the magnetic moment for Ni remains unchanged upon moving from the fcc nickel
to the compounds.

\begin{table}
\caption{Calculated local magnetic moments (in $\mu_B$)} \label{table:2} \centering
\begin{tabular}{lccl}
\hline\noalign{\smallskip}
Material & $\mu_{\mathrm{Fe}}$ & $\mu_{\mathrm{Ni}}$ & \\
\noalign{\smallskip}\hline\noalign{\smallskip}
 Ni (fcc)           &      & 0.61 & This work                           \\
                    &      & 0.62 & Exp.~\cite{Danan}                   \\
 Ni$_3$Fe (L1$_2$)  & 2.85 & 0.62 & This work                           \\
                    & 2.87 & 0.62 & Theory~\cite{Kulkova_PhysB_2002}    \\
                    & 2.97$\pm$0.15 & 0.62$\pm$0.05 & Exp.~\cite{Shull} \\
 NiFe (L1$_0$)      & 2.59 & 0.61 & This work                           \\
                    & 2.6  & 0.6  & Theory~\cite{Guenzburger}           \\
\noalign{\smallskip}\hline
\end{tabular}
\end{table}

\begin{figure}[tbp]
\centering \resizebox{0.45\textwidth}{!}{
\includegraphics{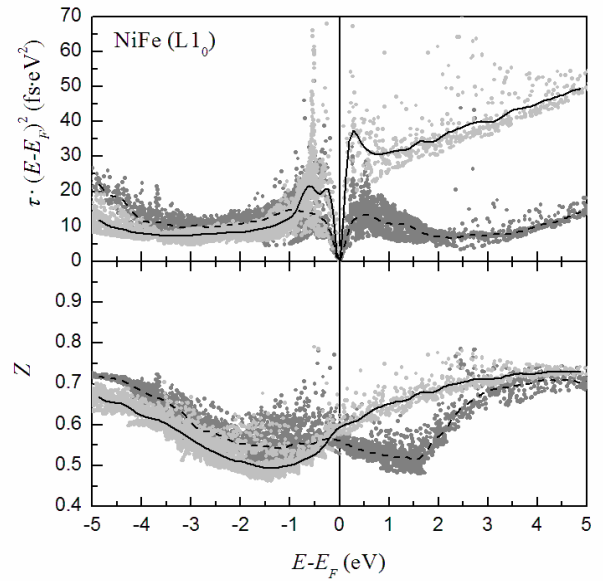}}
  \caption{Scaled lifetime $\tau_{\mathbf{k}n\sigma}\times(\epsilon_{\mathbf{k}n\sigma}-E_F)^2$ (upper panel) and
  renormalization factor $Z_{\mathbf{k}n\sigma}$ (lower panel) as functions of the exciting energy
  $\epsilon_{\mathbf{k}n\sigma}-E_F$ for spin-up (light gray points) and spin-down (dark gray points) quasiparticles
  in the L1$_0$ NiFe ferromagnetic compound. Solid and dashed lines show the corresponding scaled lifetimes
  averaged over momentum $\mathbf{k}$ for a given exciting energy.} \label{fig:7}
\end{figure}

Now, we turn to the analysis of the lifetime and the renormalization factor in the ferromagnetic compounds of
the Ni-Fe system. In Fig.~\ref{fig:7}, we demonstrate both momentum-resolved and momentum-averaged lifetime
(or, more precisely, the scaled one) and the renormalization factor for spin-up and spin-down quasiparticles
in the L1$_0$ NiFe. The main feature we would like to point out is a spread in values of the scaled lifetime
at a given exciting energy in the energy rage $\pm2$ eV, what we did not observe in the Co-Fe system. Also,
it is easily seen that the scaled lifetime of spin-down quasiparticles is quite symmetric with respect to
zero exciting energy. In addition to the spread of values, the difference between the lifetimes of spin-up
and spin-down electrons is not so large as, e.g., in the B2 CoFe. The same one can say about spin-up and
spin-down holes. In the energy range from $-1$ to $-3.5$ eV, the lifetime of spin-up holes is very close to
that of spin-down holes.

As regards the renormalization factor, for spin-up quasiparticles this quantity averaged over $\mathbf{k}$ is
closely approximated to that in the B2 CoFe, apart, maybe, from more smooth behavior as a function of
exciting energy due to the continuous $d$-band. For spin-down quasiparticles, the momentum-averaged
$Z_{\mathbf{k}n\downarrow}$ has two depressions (symmetric with respect to the Fermi level) owing to the
profound minimum in the $d$-band at $E_F$, which separates predominantly iron states (above $E_F$) and
predominantly nickel states (under $E_F$). This separation means that in the spin-down subsystem for
electrons the probability to be in iron $d$ states exceeds that to be in nickel $d$ states, while for holes
-- vice verse.

The obtained results on $\tau_{\mathbf{k}n\sigma}$ and $Z_{\mathbf{k}n\sigma}$ in the L1$_2$ Ni$_3$Fe are
shown in Fig.~\ref{fig:7}. The presented data denote that as the nickel content increases the spread in
values of $\tau_{\mathbf{k}n\sigma}$ at a given exciting energy and different momenta tends to be narrowed.
Around the Fermi level, on average the lifetime of spin-up and spin-down quasiparticles becomes shorter. In
addition to this, for holes in the energy range from $-1.5$ to $-3.5$ the momentum-averaged lifetime does not
show any noticeable dependence on spin. For electrons, the difference between the momentum-averaged
$\tau_{\mathbf{k}n\uparrow}$ and $\tau_{\mathbf{k}n\downarrow}$ is rather close to that in the compounds of
the Co-Fe system than to the difference in the L1$_0$ NiFe.

As can be expected from the DOS shape, the renormalization factor $Z_{\mathbf{k}n\uparrow}$ does not undergo
visible changes upon moving from the L1$_0$ NiFe to the L1$_2$ Ni$_3$Fe. In the spin-down subsystem,
substantial changes take place in the DOS at $(E-E_F)>E_F$: the part of the $d$-band lying right above $E_F$
has narrowed and the corresponding DOS has become higher. This DOS is characterized by the comparable
contributions of Fe and Ni. At that, under $E_F$ spin-down nickel states dominate. As a consequence, the
momentum-averaged factor $Z_{\mathbf{k}n\downarrow}$ has changed its behavior around $E_F$ (approximately
$\pm1.5$ eV), where now there is a wide minimum.

\begin{figure}[tbp]
\centering \resizebox{0.45\textwidth}{!}{
\includegraphics{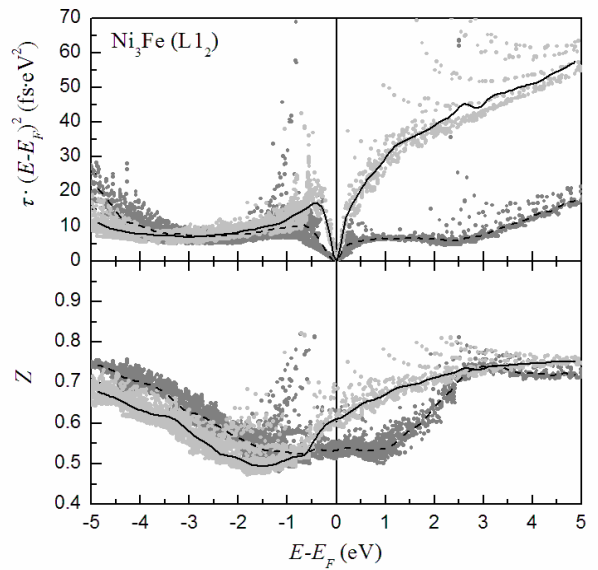}}
  \caption{Same as in Fig.~\ref{fig:7}, but in the L1$_2$ Ni$_3$Fe ferromagnetic compound.} \label{fig:8}
\end{figure}

\begin{figure}[tbp]
\centering \resizebox{0.45\textwidth}{!}{
\includegraphics{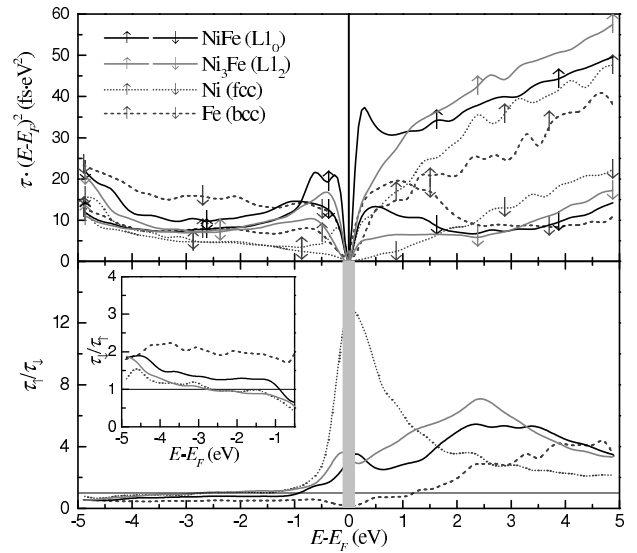}}
  \caption{Scaled lifetime averaged over momentum $\mathbf{k}$ for a given exciting energy (upper panel) and ratio
  $\tau_{\uparrow}/\tau_{\downarrow}$ between the averaged lifetimes of quasiparticles with different spin
  orientations (lower panel) as functions of the exciting energy
  for the ferromagnetic compounds NiFe and Ni$_3$Fe with the L1$_0$ and L1$_2$ structure, respectively,
  as well as for the bcc Fe and the fcc Ni. Inset in the lower panel shows
  the inverse ratio $\tau_{\downarrow}/\tau_{\uparrow}$ for holes.} \label{fig:9}
\end{figure}

In Fig.~\ref{fig:9}, we show all our results obtained for the Ni-Fe system. As is seen from the figure (see
upper panel), like in the Co-Fe system the lifetime of spin-up electrons in the compounds is longer than that
in the bcc Fe and the fcc Ni. As to the lifetime of spin-down electrons in the compounds, it demonstrates a
cross between Fe and Ni. At that, due to the aforementioned dominant contribution of iron states to the DOS
in the spin-down subsystem of the L1$_0$ NiFe, $\tau_{\mathbf{k}n\downarrow}$ in the latter is closer to that
in Fe. High atomic content of nickel in the L1$_2$ Ni$_3$Fe results in the lifetime
$\tau_{\mathbf{k}n\downarrow}$, whcih tends to be closer to that in the fcc nickel. Note that in the latter
spin-up and spin-down holes have practically the same lifetimes (except for a small vicinity of $E_F$). At
exciting energies less than $-1$ eV, the corresponding lifetimes of holes in the compounds are quite similar.

In order to analyze the spin asymmetry, in Fig.~\ref{fig:9} (lower panel) we also represent the ratio
$\tau_{\uparrow}/\tau_{\downarrow}$. Again, as in the case of the Co-Fe system, the shown data correlate with
the DOS shape and spin polarization in the vicinity of the Fermi level. As is follows from the figure, the
largest ratio corresponds to the fcc nickel, which has the highest density of spin-down states at $E_F$. The
ratio in Ni decreases from $\sim13$ to $\sim8$ away from the Fermi level to $(E-E_F)\approx0.6$ eV and
becomes smaller than $\tau_{\uparrow}/\tau_{\downarrow}$ in the compounds after $(E-E_F)\approx1.3$ eV. On
average, for electrons the L1$_2$ Ni$_3$Fe demonstrates the ratio about $5$ against $4$ in the case of the
L1$_0$ NiFe. The inset of the lower panel of Fig.~\ref{fig:9} shows the inverse lifetime
$\tau_{\downarrow}/\tau_{\uparrow}$. The presented curves reflect the situation with the lifetimes of holes,
which can be considered as a gradual transition from the bcc iron to the fcc nickel.

\begin{figure}[tbp]
\centering \resizebox{0.45\textwidth}{!}{
\includegraphics{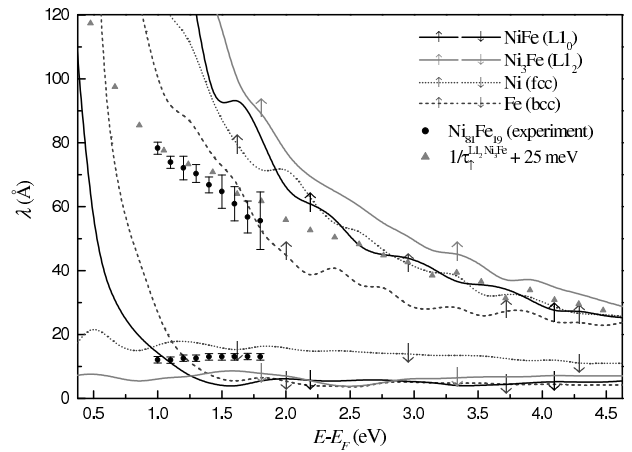}}
  \caption{Dependence of the momentum-averaged IMFP $\lambda^{e-e}_{\mathbf{k}n\sigma}$ on the exciting energy.
  The experimental data on the attenuation length for spin-up and spin-down electrons in
  Ni$_{81}$Fe$_{19}$ are taken from \cite{vanDijken}} \label{fig:10}
\end{figure}

Now, we analyze the IMFP of quasiparticles in the ferromagnetic materials considered in this subsection.
Fig.~\ref{fig:10} shows the obtained results on the IMFP of electrons as a function of exciting energy in
comparison with the experimental data on the attenuation length taken from \cite{vanDijken}, where the
spin-dependent transport of electrons in Ni$_{81}$Fe$_{19}$ films on the GaAs(001) surface has been studied.
As is clearly seen, among the materials considered here the compound Ni$_3$Fe with the L1$_2$ structure is
characterized by the largest IMFP of spin-up electrons. As to the IMFP of spin-down electrons in this
compound, it appears to be the smallest at $(E-E_F)<1.3$ eV and very close to the momentum-averaged
$\lambda^{e-e}_{\mathbf{k}n\downarrow}$ in the bcc Fe and the L1$_0$ NiFe at $(E-E_F)>2.0$ eV. Note that in
the fcc Ni and Ni$_3$Fe with L1$_2$ structure the IMFP of spin-down electrons is almost independent of
exciting energy. Such a behavior of the IMFP as a function of energy is in agreement with the experimental
data. Moreover, the calculated values are quite close to the experimental ones. However, our $G^0W^0$
calculations overestimate the attenuation length of spin-up electrons.

As well as in the case of the Co-Fe system, there is a big spin-filtering effect that is not so strong as,
e.g., in the DO$_3$ Co$_3$Fe but nevertheless disagrees with the experimental observation. As before, we
approximate the inverse value of the attenuation length of spin-up electrons in the L1$_2$ Ni$_3$Fe as
$1/\lambda^{e-e}_{\uparrow}+\Gamma_{\uparrow}/v_{\uparrow}$ but with $\Gamma_{\uparrow}=25$ meV. The obtained
results are presented in Fig.~\ref{fig:10}. We thus have achieved an agreement with the experimental
attenuation length decreasing monotonically with increasing exciting energy. This means that some
contribution that depends weakly on energy (similar to the electron-phonon broadening) should be taken into
account.

\section{Conclusions}\label{Conclusions}

In conclusion, we have presented the spin-polarized $G^0W^0$ calculations of the quasiparticle lifetime and
the quasiparticle mean free path caused by inelastic electron-electron scattering in ferromagnetic pure
metals and compounds of the Co-Fe and Ni-Fe systems. Among the compounds, we have considered CoFe and NiFe
with the B2 and L1$_0$ structure, respectively. To be closer to the Co- and Ni-rich ferromagnetic alloys used
in practice, we have also studied the DO$_3$ Co$_3$Fe and the L1$_2$ Ni$_3$Fe compounds. Such a set of
ferromagnetic materials has allowed us to demonstrate effects that structure and stoichiometry changes have
on the quasiparticle properties under study.

We have found the significant difference between the lifetime of electrons with spin up and spin down, which
is caused by the band-structure characteristic feature of the considered ferromagnetic metals and compounds.
In the DOS, this feature manifests itself as the difference between the densities of spin-up and spin-down
states in the vicinity of the Fermi energy due to the exchange splitting of the energy bands. We have
represented the resulting spin asymmetry of the lifetime by the ratio $\tau_{\uparrow}/\tau_{\downarrow}$ of
the lifetimes of spin-up and spin-down electrons. We have revealed that in the compounds the ratio can be
noticeably larger than that in their constituents. At that, owing to the originally large ratio for electrons
in the bcc Co as compared with the fcc Ni, the compounds of the Co-Fe system demonstrate
$\tau_{\uparrow}/\tau_{\downarrow}$ that on average is larger than the ratio for electrons in the compounds
of the Ni-Fe system. On the whole, with respect to the lifetime spin asymmetry the Co- and Ni-rich compounds
are preferable.

In order to estimate the contribution of the inelastic electron-electron scattering to the spin-filtering
effect experimentally observed in spin-dependent transport of electrons in the systems under study, we have
analyzed the inelastic mean free path of spin-up and spin-down electrons as a function of exciting energy. We
have shown that the mean free path expressed as the product of the lifetime and the quasiparticle velocity
inherits the lifetime spin asymmetry increased by the noticeable difference between spin-up- and
spin-down-electron velocities. As it follows from our calculations, practically within the whole positive
exciting-energy range considered in the paper the compounds of the Co-Fe system possess the longest inelastic
mean free path of spin-up electrons in comparison with the compounds of the Ni-Fe system. At that, the
inelastic mean free paths of spin-down electrons are quite similar for both systems. However, this finding
disagrees with the available experimental observations. The reason underlying such a discrepancy can relate
to decaying mechanisms additional to the inelastic electron-electron scattering. Actually, both in the Co-Fe
system and in the Ni-Fe one, we have unambiguously shown that in order to reach a satisfactory agreement with
the available experimental data on the attenuation length a contribution to the quasiparticle decay rate,
which depends weakly on exciting energy, should be taken into account in addition to the contribution coming
from the inelastic electron-electron scattering. We believe that the electron-phonon broadening contributes
significantly to the attenuation length. In order to clear up this point, spin-polarized calculations of the
electron-phonon broadening in the considered ferromagnetic compounds should be performed.

\begin{acknowledgement}
We acknowledge partial support from the Ministry of Education and Science of the Russian Federation (Grant
No. 02.740.11.5098 of 05.10.2009), the University of the Basque Country UPV/EHU (Grant No. GIC07IT36607),
the Departamento de Educaci\'on del Gobierno Vasco, and the Spanish Ministerio de Ciencia y Technolog\'ia
(MCyT) (Grant No. FIS2007-66711-C02-01). Calculations were performed on the HPC cluster of Nekrasov Kostroma
State University.
\end{acknowledgement}

\end{document}